\documentstyle[twocolumn,aps,epsf]{revtex}
\begin{document}
\title{Decoherence and the Quantum Zeno Effect}
\author{Anu Venugopalan\cite{newadd} and R. Ghosh}
\address {School of Physical Sciences,\\
Jawaharlal Nehru University,
New Delhi - 110 067, INDIA.}
\maketitle
\begin{abstract}
The measurements in the optical test of quantum Zeno effect
[Itano et al., Phys. Rev. A \underbar{41} (1990) 2295] are analyzed
using the environment-induced decoherence theory, where the
spontaneous emission lifetime of the relevant level emerges
as the `decoherence time'. The implication of this finite decoherence
time in setting a fundamental limit on the realizability of
the condition of continuous measurements is investigated in
detail.
\end{abstract}
%Fax Number $ : $ (+91)-11-686 5886\\

According to the original formulation by Misra and Sudarshan
[1], the quantum Zeno effect (QZE) refers to the inhibition
of the temporal evolution of a dynamical system under continuous
observation during a period of time. Misra and Sudarshan considered
the example of an unstable particle and showed that under 'continuous'
measurements, the particle will never be found to decay [1].
Their analysis does not take into account the actual mechanism
of the measurement process involved, but is based on an alternating
sequence of unitary time evolution and a `collapse' of the wave
function based on the projection postulate of von Neumann [2].
In performing {\it continuous} measurements, it is assumed
that successive measurements are instantaneous, yet independent
[1]. Ghirardi et al. [3] have shown that when the time-energy
uncertainty relations are taken into account, the dependence
of the measured lifetime on the frequency of measurements, although
present in principle, would be extremely difficult to observe
in the case of spontaneous decay. The QZE has not yet been observed
experimentally for spontaneous decay.

Following an original proposal by Cook [4], Itano et al. [5]
demonstrated experimentally the occurrence of the QZE as the
inhibition of {\it induced} radio frequency transitions
between two hyperfine levels of a berillium ion caused by frequent
measurements of the level population using optical pulses (see
Fig. 1). This test of QZE does not involve observation of spontaneous
decay of naturally unstable quantum system, as in the original
idea by Misra and Sudarshan. The system is initially prepared
in level $1$. The $ 1 \rightarrow 2 $ transition is driven by a resonant
radio-frequency $ \pi $ pulse. In order to observe the level
populations, level $ 1 $ is connected by an optical transition to
an additional level $ 3 $ such that level $ 3 $ can decay only to level
$1$. Spontaneous decay from level $ 2 $ to level $ 1 $ is negligible.
The measurements are carried out (during the evolution under
the $ \pi $ pulse) by driving the $ 1 \rightarrow 3 $ transition
with $ N $ equispaced short optical pulses and observing the presence
(or absence) of spontaneously emitted photons from level $ 3 $ corresponding
to the atom being found in the level $ 1 $ (or $2$). In the experiment
in ref. [5], a `freezing' of population in one level was observed
as a result of frequent measurements. There have been many attempts
to theoretically explain this experimental observation of the
QZE [5-7], and to classify the various examples of quantum Zeno
effect (or `paradox') in terms of either strictly unitary evolutions
or `wave-function collapse' by measurement.

\bigskip
\epsfxsize=3.0in
\epsfysize=3.0in
%\centerline{\epsfbox{zeno-fig.eps}}
\medskip
{\raggedright Figure 1 {\it
Energy-level diagram for the quantum Zeno
effect experiment [5]. In our model, the environment enters
as the collection of vacuum modes coupled to level $3$.}}
\bigskip

In this Letter we apply the environment-induced decoherence
theory [8] to analyze the observed results in the QZE experiment
[5], and provide connection with the theoretical analysis in
ref. [6]. We demonstrate that a fundamental limit on the frequency
of measurements emerges from the theory, which is in tune with
the predictions in ref. [3], and can be contrasted with the
recent findings in ref. [7] in the context of QZE with neutron
spin.

By using the postulate of projection of the wave function [2]
Itano et al. have shown that in the limit of infinitely frequent
measurements, the probability of one of the levels being populated
goes to unity in the following way [5]. Suppose the ion is in
level $ 1 $ at time $ \tau = $ 0. An on-resonance $\pi$-pulse
of duration $ T = \pi /\Omega $ is applied, where $ \Omega $ is the
Rabi frequency. Without the measurement pulses,
the probability $ P_{2}(T)$ for the ion to be in level
$ 2 $ at $ \tau = T $ is $1$. Let $ N $ measurement pulses be applied
within $ T $ at times $ \tau_{k}= kT/N$, $ k = 1,
2, ..., N$. The level populations at the end of the $ \pi $ pulse
are calculated using the Bloch vector representation of a two-level
system [5]. In the rotating wave approximation, the system can
be described by a Bloch vector ${\bf R} \equiv (R_{1 },R_{2 }, R_{3 })$,
whose components interms of the density matrix elements are
\begin{eqnarray}
 R_{1} &\equiv& \rho_{12 }+ \rho_{21 }, \nonumber\\
 R_{2} &\equiv& i (\rho_{12 }- \rho_{21 }) ,\nonumber\\
 R_{3} &\equiv& \rho_{22 }- \rho_{11 }\equiv P_{2 }- P_{1},
\end{eqnarray}
where $ P_{1 }(t)$ and $P_{2 }(t)$ are the probabilities for the ion
to be in the levels $ 1 $ and $2$, respectively. The equation of motion
for {\bf R} is
\begin{equation}
 { d{\bf R}\over dt} = \omega X {\bf R } , \eqnum{2}
\end{equation}
where $ \omega = (\Omega $ ,0,0). The {\it projection }{\it postulate}
for each measurement essentially amounts to setting the coherences
$ (\rho_{12}, \rho_{21 })$ to zero while leaving the populations
$ (\rho_{11 }, \rho_{22 }) $ unchanged. It can be easily
seen [5] that after $ N $ measurements (projections) the population
$ P_{2}(T) = [R_{3 }(T) + 1]/2$ at the end of the $ \pi $ pulse is
\begin{equation}
 P_{2 }(T) = [1 - cos^{N }(\pi /N)]/2. \eqnum{3}
\end{equation}
For large $N$,
\begin{eqnarray}
 P_{2 }(T) &\cong& \lim_{N\to \infty}[1 - (1 - \pi^{2 }/2N^{2 })^{N }]/2
 \nonumber\\
&\cong& \lim_{N\to \infty}[1- \exp(- \pi^{2 }/2N)]/2 \cong 0. \eqnum{4}
\end{eqnarray}
As $ N \rightarrow \infty$, $P_{2 }(T)$ decreases monotonically
to zero. This, according to Itano et al. [5], is the explanation
of the observed Zeno effect.

In another approach, Frerichs and Schenzle [6] have shown that
the outcome of the experiment in ref. [5] can be explained by
looking at the optical Bloch equations for the complete
{\it three}-level system. These semiclassical equations of motion for
the elements of the density matrix take into account the dissipation
effects due to spontaneous emission along the optical transition path.
Under the condition that the spacing $T/N$ of the measurement
pulses is greater than the spontaneous emission lifetime of
the third level, it has been demonstrated by numerical integration
of the Bloch equations [6] that all measurable results of quantum
Zeno experiment emerge naturally from continuous irreversible
quantum dynamics, without having to appeal to concepts such
as the projection postulate or the wave function collapse. A
number of recent studies attempt to explain the observed inhibition
of transitions using the unitary evolution given by the Schrodinger
equation, but provide no specific mechanism for the measurement
process.

We propose that the measurement processes involved in the experiment
of Itano et al. [5] can be explained by the environment-induced
decoherence approach [8]. This approach is based on the understanding
that a macroscopic apparatus is never isolated from its environment.
Note that the actual measurement process in the QZE experiment
[5] involves the observation (or non-observation) of {\it spontaneously}
emitted photons from level $ 3 $ (see Fig.1). While the irreversible
processes of spontaneous emission are accounted for by phenomenological
decay rates in the semiclassical theory, a fully quantized theory
of the field and atom is required to explain it. It is well
known in the Weisskopf-Wigner theory [9] that spontaneous emission
decay rates emerge naturally when a completely quantized field
treatment including the coupling to the field vacuum modes,
i.e., {\it the environment}, is considered. Since one is
interested only in the dynamics of the atom, an average over
the large number of degrees of freedom of the multimode vacuum
leads to an irreversible dynamic process. The reduced density
matrix of the atom is, thus, driven to a diagonal form over
a time scale equal to the spontaneous emission lifetime of the
excited level. This treatment is exactly that of the `environment-
induced decoherence' approach [8]. The experiment of Itano et al. [5]
can now be viewed in the following way. The two-level system
(levels $ 1 $ and $2$) constitutes the `system' of interest for which
QZE is to be investigated. Level $ 3 $ is coupled to the system
of interest in such a way that it can be used to measure the
level populations (i.e., to determine whether the ion is in
level $ 1 $ or 2) and hence level $ 3 $ can be interpreted as the
`apparatus'. The collection of vacuum modes is the `environment' the
effect of which is significant only for level $ 3 $ (see Fig. 1) as it
has a finite spontaneous emission lifetime. Since level $ 2 $ is
assumed to be metastable, the system is coupled only to the
apparatus and is isolated from the environment. The system-apparatus
interaction is through the short optical pulses that connect
levels $ 1 $ and $3$. The equation for the reduced `system-apparatus'
composite after tracing over the environment variables are the
optical Bloch equations considered by Frerichs and Schenzle
[6]. The interaction with the environment drives the density
matrix of the apparatus to a diagonal form. Since the apparatus
is in turn coupled to the system, the density matrix of the
system-apparatus composite acquires a diagonal form. The environment-
induced decoherence approach provides a mechanism for {\it deriving}
the phenomenological damping in the Bloch equations in the way
described above.

The physical meaning and realizability of the `continuous measurements'
limit as incorporated in the example of spontaneous decay of
an unstable particle in the original formulation of the Zeno
problem was criticised by Ghirardi et al. [3] using arguments
based on the time-energy uncertainty relations. In the following
we show that in the environment-induced decoherence theory,
there exists a fundamental limit on the maximum number of measurements
that can be performed in a set-up similar to that in ref. [5].
The crucial point here is that the {\it decoherence } of
the reduced density-matrix of the system-apparatus (when the
density matrix collapses to a diagonal form) does not take place
instantaneously, but over a characteristic time scale - the
{\it decoherence time}. It is obvious that the decoherence
time for each measurement is the spontaneous emission lifetime
of the level $3$. This sets a fundamental limit on the requirement
of `continuous measurements' for the QZE since the photons from
level $ 3 $ cannot be observed at a rate faster than the decoherence
rate. The limit of infinitely frequent or continuous measurements
is then a mathematical idealization. Let us consider the analysis
of Itano et al. again. The measurements are made at times $ \tau_{k }=
kT/N$, $ k = 1, 2, ..., N$, where for a $ \pi$-pulse,
$ T = \pi /\Omega$, $\Omega $ being the Rabi frequency. The probability,
$ P_{2 }(T)$, of the ion being in level $ 2 $ as given by (3) monotonically
decreases to zero as $ N \rightarrow \infty$. However, since the time
between the measurements is limited by the spontaneous emission lifetime
$ \tau_{sp}$ of the third level, there exists a {\it lower bound} on the
argument of the cosine term in (3), viz.,
\begin{equation}
 \pi /N \ge \Omega \tau_{sp }. \eqnum{5}
\end{equation}
Therefore, for large N, the expression (4) for $ P_{2}(T)$ becomes
\begin{eqnarray}
 P_{2 }(T) &\cong& \lim_{N\to \infty}
[1 - (1 - \Omega^{2}\tau^{2}_{sp }/2)^{N }]/2 \nonumber\\
&\cong& \lim_{N\to \infty} [1 - \exp(- \Omega^{2}
\tau^{2}_{sp}N/2)] /2. \eqnum{6}
\end{eqnarray}

In the limit of $ N \rightarrow \infty$, the above expression {\it
does} not tend to zero, but approaches one-half .{\it } As
explained earlier, the theoretical analysis of Itano et al.
assumes the artificial collapse hypothesis with no dynamical
mechanism for the measurement, and Eq.(3) is said to be the
explanation of the observed effect in the limit of $N\rightarrow\infty$.
If one tries to incorporate the finite measurement
time into the analysis of Itano et al., it leads to the paradoxical
situation described by Eq.(6). From (5) it is obvious that $N$ cannot
increase indefinitely to infinity but is limited to $N_{max}$, where
\begin{equation}
 N_{max }\equiv \pi / (\Omega \tau_{sp }) = T/\tau_{sp }. \eqnum{7}
\end{equation}
The origin of this limitation on the maximum value of $ N $ is the
finite spontaneous emission lifetime of the level $3$, and hence
can be traced back to the time-energy uncertainty relation [3].

In a real experiment, increasing the number of measurement pulses
$ N $ beyond $ N_{max}$ would amount to making
the $ 1 \rightarrow 3 $ transition almost continuous. Since a measurement
is defined solely by the observation or non-observation of spontaneously
emitted photons from level 3, the process of measurement of
level population would be ill-defined in such a situation since
level $ 3 $ would not be allowed to decay to level $1$. For
$ N > N_{max}$, the population gets stuck in level $3$, and the
non-observation of photons from level $ 3 $ here does not imply that
the population is stuck in level $ 2 $ as is the case in QZE. Thus
increasing $ N $ beyond $ N_{max}$ is unsuited from the point of view
of the QZE.

Recently, Nakazato et al. [7] have studied the QZE in the case
of neutron spins which is similar in spirit to the system studied
by Itano et al. [5]. Nakazato et al. have shown that the limit
of continuous measurements is unphysical. In their example,
if the neutron is initially prepared in the spin-up ($\uparrow$)
state in the presence of a magnetic field, it evolves to the
spin-down state after a time $ T $ in a way similar to the two-level
ion going from level $ 1 $ to level $ 2 $ on the application of a
$\pi$ pulse in Ref. [5]. Now, if one does $ N $ measurements of the
state at regular time-intervals of $T/N$ during $T$, the probability
that the neutron spin is up at time $ T $ in the limit $ N \rightarrow
\infty$ is
\begin{equation}
 P_{\uparrow}(T) = \lim_{N\to \infty} [cos^{2 }(\pi /2N)]^{N }\cong 1.
 \eqnum{8}
\end{equation}
Here $ \pi $ /2N $ = \mu $ B{\it l} $ /\hbar v$,
where $ B $ is the applied static magnetic field, $ \mu $ is
the modulus of the neutron magnetic moment, {\it l} is the
length of the region where $ B $ is present, v{\it } is the
neutron speed and $ \hbar $ is Planck's constant. Nakazato
et al. [7] argue that from a physical point of view it is impossible
to avoid uncertainties in the neutron speed $ \Delta v$
and the position $ \Delta x$. They show that the argument
$ \phi \equiv \pi /2N$ of the cosine term has a lower
bound, $ \phi_{0 }= \Delta E_{m }/4\Delta E_{k }, $ where
$ \Delta E_{m }= 2\mu B$ is the magnetic energy gap, and
$ \Delta E_{k }=\Delta (mv^{2 }/2)_{v=v_{0} }$ is the spread
in the kinetic energy of the neutron beam at the mean speed
$ v_{0 }. $ Thus, for large $N$,
\begin{eqnarray}
P_{\uparrow}(T) &\cong& (cos\phi_{0 })^{2 N }
\cong (1 - \phi_{0}^{2}/2)^{2N}\cong exp(- \phi_{0}^{2}N). \eqnum{9}
\end{eqnarray}
Thus, as $ N\rightarrow\infty$ , $P_{\uparrow}(T)$ {\it vanishes}
{\it to }{\it zero } unlike in (8). Nakazato et al. thus conclude
that the limit $ N \rightarrow \infty $ is unphysical as it
leads to the paradoxical result (9) when the uncertainty relations
are taken into account. Nakazato et al. [7] propose a limiting
value of $ N $ which is obtained by setting $ P_{\uparrow }(T)
\cong 1/2$ in (9). Note that if we deduce the maximum value
of $ N $ from the lower bound $ \phi_{0 }$ of $ \pi /2N$,
so that
\begin{equation}
 N_{max }= \pi /(2\phi_{0 })= 2\pi \Delta E_{k}/\Delta E_{m },
 \eqnum{10}
\end{equation}
then by a proper choice of $ N_{max }$, $P_{\uparrow }(T)$ can be
made to exceed 1/2, and it can be very close to unity. Thus the
assertion of $ P_{\uparrow}(T) \cong 1/2$ at $ N \cong N_{max}$
in [7] is not quite justified.

In summary, we have shown that the {\it irreversibility}
in the quantum dynamical equations associated with the
{\it measurements} in a quantum Zeno experiment [5] can be derived
using the environment-induced decoherence approach in which the
interaction of an atomic level with the vacuum modes of the quantized
field results in spontaneous emission from the atomic level. In the
experiment, the observation of spontaneously emitted photons from a
level defines a measurement. The finite time taken by the atomic
density matrix to {\it decohere} (and hence yield a measurement) under
such an interaction with the vacuum modes is the spontaneous emission
life-time of the atomic state concerned. This finite decoherence time
limits the frequency of measurements in the quantum Zeno set-up. Our
results are compared with those in the recently studied example of
quantum Zeno effect in neutron spins [7].

RG acknowledges partial support from a grant by the Department
of Science and Technology, Government of India.

\end{document}